\documentclass[a4paper,11pt]{article}
\pdfoutput=1

\usepackage{jinstpub} 

\usepackage{todonotes}

\usepackage{comment}
\usepackage{cleveref}

\title{ZynqMP-based board-management mezzanines for Serenity ATCA-blades}

\author[a]{T. Mehner\note{Corresponding author.},}
\author[a]{L. E. Ardila-Perez,}
\author[a]{M. N. Balzer,}
\author[a]{O. Sander,}
\author[a]{D. Tcherniakhovski,}
\author[a]{M. Schleicher,}
\author[a]{M. Fuchs,}
\author[c]{G. Fedi,}
\author[c]{G. Gimas,}
\author[c]{G. M. Iles,}
\author[c]{M. Pesaresi,}
\author[c]{A. W. Rose}
\author[b]{and T. Schuh}

\affiliation[a]{Karlsruhe Institute of Technology, Institute for Data Processing and Electronics (IPE), \\Hermann-von-Helmholtz-Platz 1, D-76344 Eggenstein-Leopoldshafen, Germany}
\affiliation[b]{STFC Rutherford Appleton Laboratory, Particle Physics Department, \\Harwell Campus, Didcot, OX11 0QX, UK}
\affiliation[c]{Imperial College London, Physics Department, Blackett Laboratory, \\Prince Consort Rd, London, SW7 2BW, UK}

\emailAdd{torben.mehner@kit.edu}

\abstract{In the context of the CMS Phase-2 tracker back-end processing system, two mezzanines based on the Zynq Ultrascale+ Multi-Processor System-on-Chip (MPSoC) device have been developed to serve as centralized slow control and board management solution for the Serenity-family \textcolor{black}{Advanced Telecommunications Computing Architecture (ATCA)} blades.

This paper presents the developments of the MPSoC mezzanines to execute the Intelligent Platform Management Controller (IPMC) software in the real-time capable processors of the MPSoC. In coordination with the Shelf Manager, once full-power is enabled, a CentOS-based Linux distribution is executed in the application processors of the MPSoC, on which EMPButler and the Serenity Management Shell (SMASH) are running.
}

\keywords{Detector control systems (detector and experiment monitoring and slow-control systems, architecture, hardware, algorithms, databases), Digital electronic circuits, Trigger concepts and systems (hardware and software)}

\collaboration[c]{on behalf of the Serenity collaboration}

\proceeding{Topical Workshop on Electronics for Particle Physics - TWEPP2021\\
  on 20 -- 24 September, 2021\\
  Online Conference}

\begin{document}
\maketitle
\flushbottom

\section{Introduction}

One of the enhancements in context of the high-luminosity upgrade of the LHC (HL-LHC) is executing a track reconstruction based on a time-multiplexed architecture requiring the use of two layers of data processing~\cite{Contardo, HALL2016292}. The first layer is represented by the Outer Tracker Data, Trigger, and Control (OT DTC) boards, which extract and pre-process the pre-selected detector data and route it to the corresponding card in the Track Finding layer. The DTC cards are also responsible for sending the raw hit data to the DAQ \& TTC Hub (DTH) cards and providing the control and timing signals to the tracker front-end modules. The second layer, the Track Finding Processor (TFP), receives all pre-selected detector data for a given event and reconstructs the trajectories of charged particles in the tracker.

The Serenity ATCA board family is a result of the R\&D program currently underway within the level-1 trigger community. The boards are candidates for the OT DTC and will potentially also be used in other applications~\cite{Collaboration:2283192}. Among the technologies being explored with Serenity are ZynqMP based slow control modules that enable a tight integration of the intelligent platform management controller (IPMC) and a slow control CentOS based Linux in a single system on chip (SoC) device. The architecture of this approach is presented in the following.

\section{Serenity ATCA-blades}

\begin{figure}[h]
\centering
\includegraphics[width=.4\linewidth]{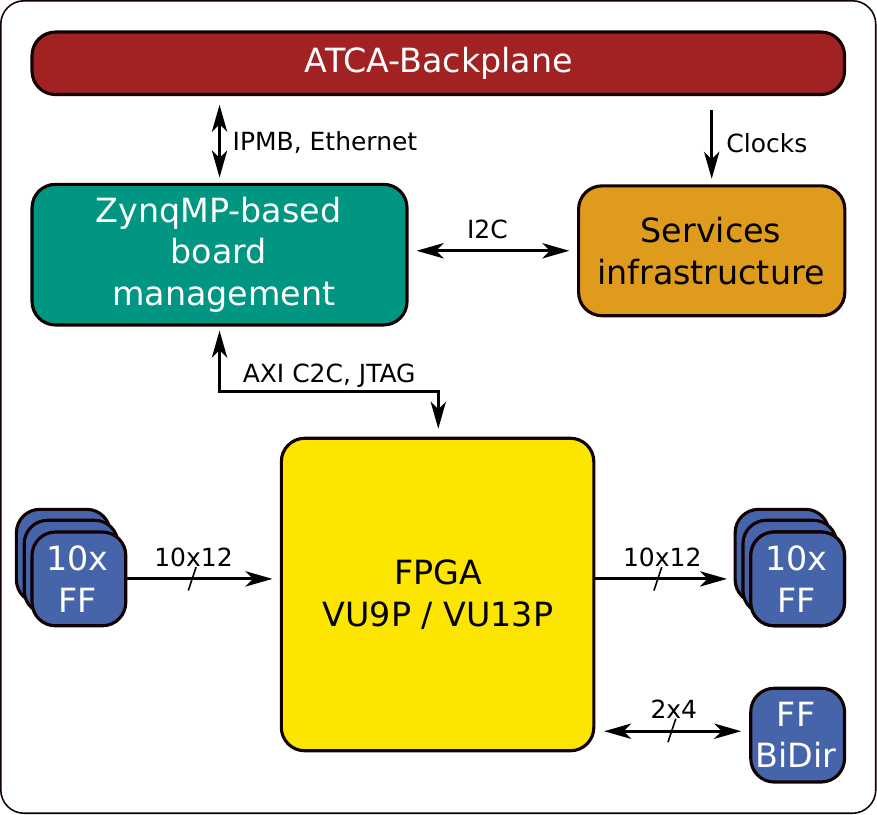} \qquad
\includegraphics[width=.35\linewidth]{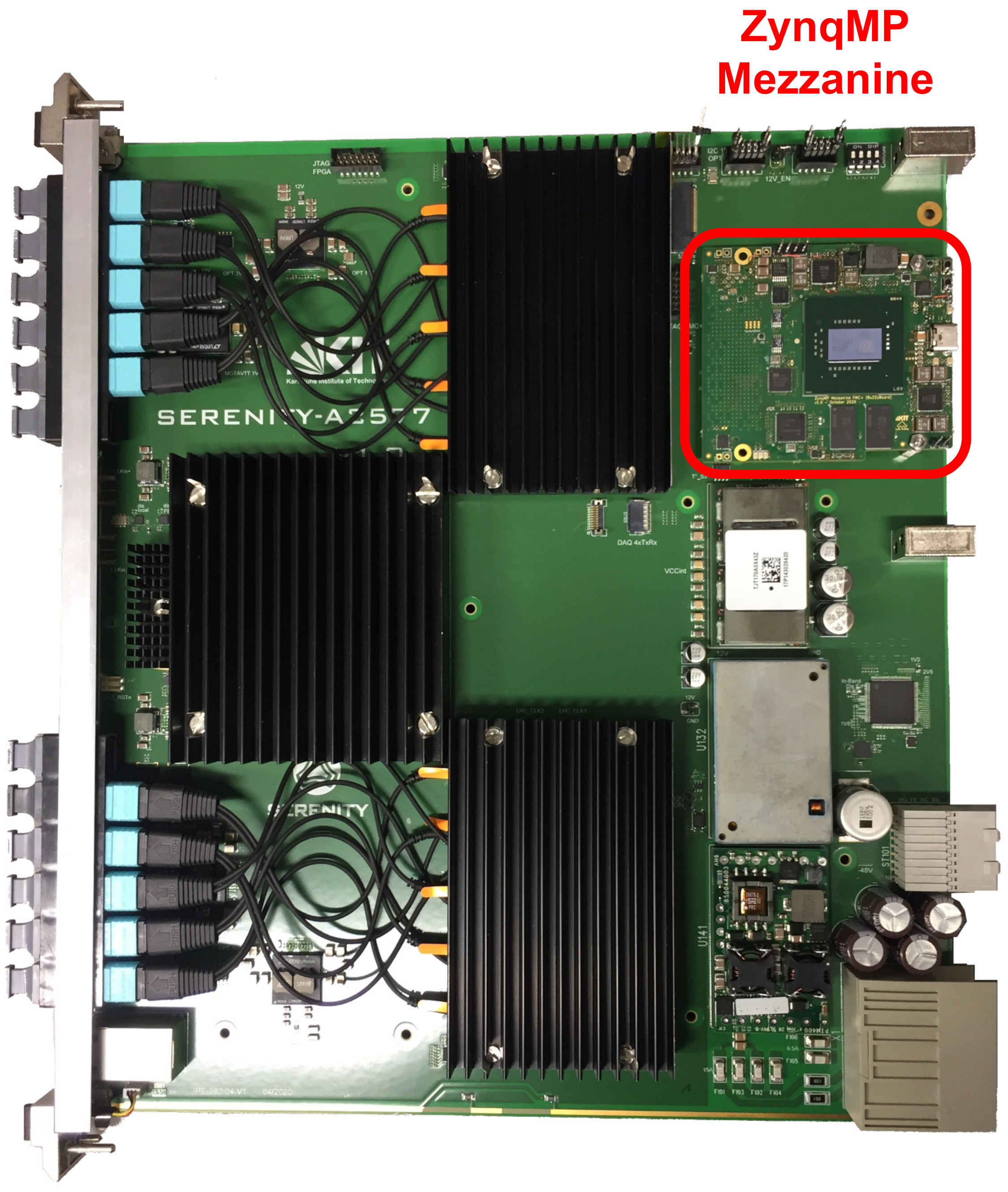}
\caption{Serenity-A Architecture \textcolor{black}{(left)} and Board \textcolor{black}{(right)}}
\label{fig:serenity-a-architecture}
\end{figure}

The Serenity family consists of the two ATCA-blades Serenity-A and Serenity-Z. The blades are centered around high-end processing FPGAs that are connected to Samtec FireFly optical high-speed modules for data transfer. Serenity-A is centered around a single Virtex Ultrascale+ VU9P or VU13P FPGA. Serenity-Z provides two Samtec Z-Ray interposers for FPGA daughtercards. These daughtercards can either contain Kintex Ultrascale+ KU15P or Virtex Ultrascale+ VU7P, VU9P or VU13P FPGAs.

Besides the regular ATCA board infrastructure like power, clock and IPMC both boards utilize a slow-control processor for control, calibration, and monitoring of the blade thereby adding significant computing power to each blade. Serenity-Z uses an x86 based module, an IPMC Dimm module, and a glue-logic ARTIX FPGA~\cite{a}. For the connection of the board-management mezzanine, the Serenity-Z has an extended ComExpress (CMX-EXT) connector, which is a standard ComExpress connector with an auxiliary connector outside the ComExpress-footprint. Serenity-A explores an alternative approach that integrates these functionalities into one single Xilinx ZynqMP FPGA. For attaching the board-management mezzanine, it provides a custom FMC+ connector. Serenity-Z can also accommodate a ZynqMP mezzanine with an extended ComExpress interface. An overview of Serenity-A is depicted in Fig.\,\ref{fig:serenity-a-architecture}.

\section{ZynqMP-based board-management mezzanines}
\label{sec:zynqmp-board-management-mezzanines}

As there is no commercial \textcolor{black}{System-on-Module} with a Zynq Ultrascale+ Multi-Processor System-on-Chip (MPSoC) available that exposes all of its high-speed lanes to the connector, the ZynqMP-based board-management mezzanine was developed. A Zynq Ultrascale+ 4EG MPSoC, that provides an application processor, a real-time processor and a \textcolor{black}{P}rogrammable \textcolor{black}{L}ogic (PL), has been selected due to it being the smallest Zynq MPSoC still having high-speed lanes.

\begin{figure}[h]
\centering
\includegraphics[width=.7\linewidth]{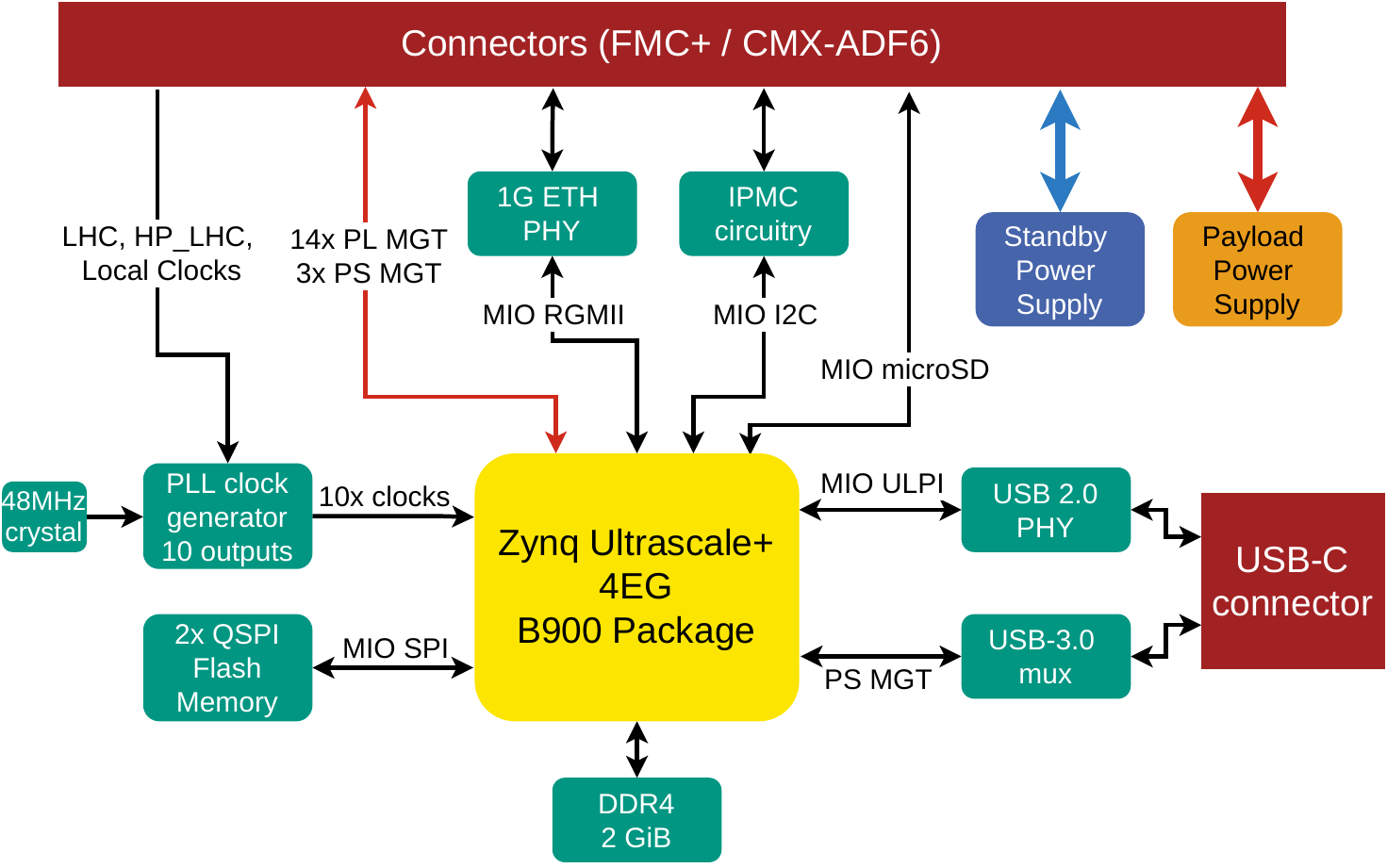}
\qquad
\includegraphics[width=.225\linewidth]{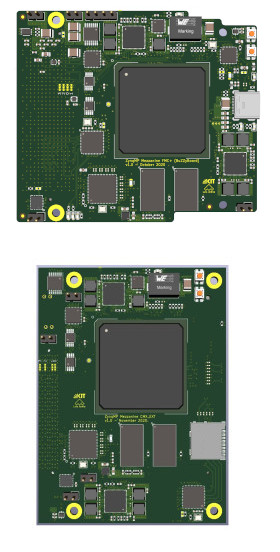}
\caption{ZynqMP Mezzanine Architecture \textcolor{black}{(left), FMC+ Board (top right) and CMX-EXT Board (bottom right)}}
\label{fig:zynqmp-mezzanine}
\end{figure}

The mezzanine is available in FMC+ and CMX-EXT form-factor to be used with \textcolor{black}{both Serenity-A} and Serenity-Z. \textcolor{black}{Despite} having different form-factors, the part of the mezzanine around the MPSoC and RAM is identical to both boards. To communicate with the shelf-manager, the mezzanine provides the necessary connections to the backplane. Furthermore, flexible high-speed links and low-speed connections as well as an Ethernet-connection are available to the ATCA blade. The MPSoC can boot from either an on-board QSPI flash or an external microSD-card on the ATCA-blade. As the ATCA standard limits standby-power draw before negotiating a switch to payload-power, there are two independent power supplies for the \textcolor{black}{Low-Power Domain} (LPD) and \textcolor{black}{Full-Power Domain} (FPD) including PL of the ZynqMP. Starting with only the LPD supplied, the mezzanine enables FPD and PL when payload power is activated. All of this was put on each 14-layer small-footprint board, which are shown in Fig.\,\ref{fig:zynqmp-mezzanine}.

\section{Software and Firmware Framework}

\begin{figure}[h]
\centering
\includegraphics[width=\linewidth]{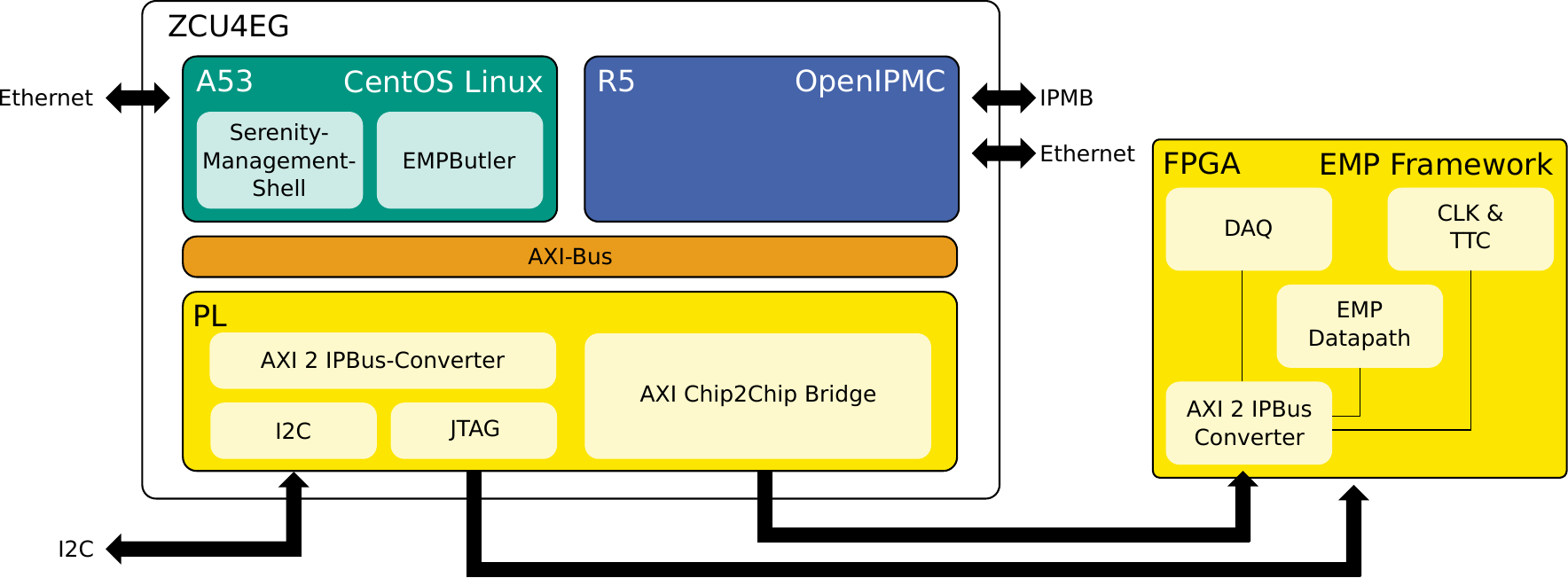}
\caption{Software and Firmware Architecture}
\label{fig:sowftware-and-firmware-architecture}
\end{figure}

The following description of software and firmware focuses on the ZynqMP, but being mostly hardware independent, this framework is also used in x86 systems with only minor adjustments. As Fig.\,\ref{fig:sowftware-and-firmware-architecture} shows the ZynqMP incorporates the IPMC, the Linux-based board management and glue-logic in the device. While the IPMC is in charge of the communication between ATCA-blade and shelf manager, the Linux-based applications are responsible for slow control. Thereby the glue-logic adapts and extends the interfaces of the \textcolor{black}{Processing System (PS)} part of the ZynqMP. The figure further shows a processing FPGA incorporating the EMP framework. This framework enables processing of high-speed data through an interchangeable algorithm~\cite{d}.

\paragraph{Integrated OpenIPMC}

The IPMC is running on the ARM-R5 \textcolor{black}{Real-Time Processing Unit} (RPU) of the ZynqMP. It is possible to use either the well-known commercial Pigeon Point IPMC software or the open-source OpenIPMC-software. Both versions have been successfully integrated and tested on the Zynq Ultrascale+~\cite{b, c}\textcolor{black}{.}

On startup, the ZynqMP runs off standby-power until the IPMC negotiates the transition to payload-power with the shelf manager. Only the LPD of the ZynqMP, to which the RPU belongs, is powered from the beginning, therefore only the IPMC is running. To stay in the limited ATCA standby power budget Linux and PL are not configured at this stage. Only after negotiating the power state, the rest of the ZynqMP is switched on. Then the FPGA is configured and Linux is started.

\paragraph{CentOS Linux, SMASH and EMPButler}

The CentOS-based Linux runs on the ARM A53 \textcolor{black}{Application Processing Unit} (APU) of the ZynqMP. It consists of a PetaLinux-kernel with a CentOS root-filesystem. EMPButler and Serenity Management Shell (SMASH) are the applications that run within the Linux. EMPButler is responsible for controlling the EMP framework in the processing FPGAs on the ATCA-blade. SMASH controls the services infrastructure on the ATCA-blade and is able to program the processing FPGA.

Both of these applications use the IPBus-protocol~\cite{e} to connect to its endpoints. The memory-mapped interface of IPBus is used for connecting IPBus endpoints through AXI and AXI-C2C. The interface is added to Linux as a uio-device that allows only the specific region of memory that belongs to the IPBus interface to be read and written and can therefore be used by non-root users.

The EMPButler's IPBus endpoints are registers within the EMP-framework that is used on the processing FPGA. To connect to these registers, the AXI-to-IPBus-converter is instantiated in the processing FPGA itself, with the AXI being extended over an AXI Chip-2-Chip-Bridge.

SMASH also relies on IPBus and is used to configure various board components. For example, SMASH can program the clock chip or read the optical module's supply voltage via I2C. It furthermore can program the processing FPGA using JTAG. Accordingly, SMASH's IPBus endpoints are I2C and JTAG bus masters. These are, together with the AXI-to-IPBus converter instantiated in the MPSoC's PL.

\paragraph{ZynqMP split boot process}

Assuming hundreds of boards in the back-end electronics, software update mechanisms become a concern. Booting up-to-date images via network (e.g. PXE based) is a standard in IT for many years and extends to x86 ComExpress modules. Yet, fetching the whole configuration via network on boot is not supported out of the box for ZynqMP.

The ZynqMP already allows to fetch most of the firmware and software at run-time from a network source as early as the U-Boot stage (kernel, bitstream, rootfs). However, the PS configuration can not be loaded via network but is application design specific. To be able to fetch all application specific data at run time a two-stage configuration process of the PS has been developed which implements a generic first stage and an application specific second stage. For most of the additional steps, tools and scripts were designed~\cite{f} to handle the two-stage configuration of the PS as well as other necessary modifications to adapt the boot process to the ATCA standard.

The proposed boot process starts with executing the \textcolor{black}{First-Stage Bootloader} (FSBL), that among others applies the first stage configuration to the PS. It also loads the IPMC software for execution on the RPU as well as U-Boot as a \textcolor{black}{Second-Stage Bootloader} (SSBL) for execution on the APU. The second-stage bootloader is now capable to load data from a network source. Thus it is used to replace the initial configuration of the PS with the second stage one and also to fetch a bitstream from the network and load it to the PL. Afterwards U-Boot loads the Linux kernel likewise and starts it. During booting, the kernel then mounts a root-filesystem using NFS. While this boot process uses proven concepts wherever possible, like the implementation of PXE in U-Boot to load the kernel and NFS for the root-filesystem, it further adds novel, previously missing links, to make the highly integrated ZynqMP-based board-management mezzanines a competitive alternative to solutions based on off the shelf x86 ComExpress modules.

\section{Conclusion}

For the Serenity ATCA blades that will be used in the CMS phase-2 tracker installation, two ZynqMP-based board-management mezzanines were developed. These mezzanines are available in FMC+ and CMX-EXT form-factor.

On the RPU of the \textcolor{black}{ZynqMP}, an IPMC software is running. The IPMC negotiates with the shelf-manager to switch to payload-power, before enabling power to the rest of the ZynqMP.

A CentOS-based Linux is then running on the APU of the ZynqMP. Within Linux, EMPButler, which is controlling the EMP framework on the processing FPGA and SMASH, which is controlling the services infrastructure are running. These applications need an AXI-to-IPBus converter in the PL to connect to their respective endpoints, which for SMASH are also instantiated in the PL.

Using a split boot process, it is possible to fetch all application specific configuration, including the PS configuration, from a network storage.


\begin{thebibliography}{99}

\bibitem{Contardo}
D. Contardo et. al., \emph{Technical Proposal for the Phase-II Upgrade of the CMS Detector}, CERN-LHCC-2015-010, (2015)

\bibitem{HALL2016292}
G. Hall, \emph{A time-multiplexed track-trigger for the CMS HL-LHC upgrade}, Nucl. Instrum. Methods Phys. Res. 824, p. 292 -- 295, (2016)

\bibitem{Collaboration:2283192}
CMS Collaboration, \emph{The Phase-2 Upgrade of the CMS L1 Trigger Interim Technical Design Report}, CERN-LHCC-2017-013, (2017)

\bibitem{a}
A. Rose et. al., \emph{Serenity: An ATCA prototyping platform for CMS Phase-2}, Topical Workshop on Electronics for Particle Physics (TWEPP2018), (2018).



\bibitem{b}
\textcolor{black}{L. Ardila-Perez et. al., \emph{A novel centralized slow control and board management solution for ATCA blades based on the Zynq Ultrascale+ System-on-Chip}, EPJ Web Conf. 245 01015, (2020)}

\bibitem{c}
\textcolor{black}{L. Calligaris et al., \emph{OpenIPMC: A Free and Open-Source Intelligent Platform Management Controller Software}, IEEE Transactions on Nuclear Science, vol. 68, no. 8, pp. 2105-2112, (2021)}

\bibitem{d}
A. Thea et. al., \emph{Common infrastructure F/W for Phase 2 hardware},
Workshop on CMS Tracker BackEnd Systems \& DAQ for Phase-2, (2018)

\bibitem{e}
C. Ghabrous Larrea et. al., \emph{IPBus: a flexible Ethernet-based control system for xTCA hardware}, JINST 10, (2015).

\bibitem{f}
L. Ardila et. al., \emph{ZynqMP-based board-management mezzanines for the Serenity ATCA-blades}, CERN System On Chip Workshop, (2021).

\end{thebibliography}
\end{document}